\documentclass[aps,prl,twocolumn,showpacs%
]{revtex4}
\usepackage[hyperindex,breaklinks]{hyperref}

\begin{document}
\def\dd{{\rm d}} \def\ds{\dd s} \def\e{{\rm e}} \def\etal{{\em et al}.}
\def\al{\alpha}\def\be{\beta}\def\ga{\gamma}\def\de{\delta}\def\ep{\epsilon}
\def\et{\eta}\def\th{\theta}\def\ph{\phi}\def\rh{\rho}\def\si{\sigma}

\def\mean#1{{\vphantom{\tilde#1}\bar#1}}
\def\bH{\mean H}\def\OM{\mean\Omega}\def\ab{\mean a} \def\bz{\mean z}
\def\rhb{\mean\rh}\def\bq{\mean q}\def\bT{\mean T}\def\bn{\mean n}
\def\bD{\mean D}\def\mx{\mean x} \def\rb{\mean r}
\def\OmB{\Omega\Z B}\def\bOmB{\mean\OmB}\def\bnB{\bn\Z B}
\def\gb{\mean\ga}\def\gc{\gb\Z0}\def\etb{\mean\eta}

\def\w#1{\,\hbox{#1}} \def\Deriv#1#2#3{{#1#3\over#1#2}}
\def\Der#1#2{{#1\hphantom{#2}\over#1#2}} \def\br{\hfill\break}
\def\ts{t} \def\tc{\tau} \def\Dts{\mathop{\hbox{$\Der\dd\ts$}}}
\def\Dtc{\mathop{\hbox{$\Der\dd\tc$}}} \def\tb{\ts'}
\def\ns#1{_{\hbox{\sevenrm #1}}} \def\dOM{\dd\Omega^2}
\def\goesas{\mathop{\sim}\limits} \def\tn{\ts\Z0}
\def\zb{{\bar z}} \def\bxi{{\mathbf\xi}}
\def\Y#1{^{\raise2pt\hbox{$\scriptstyle#1$}}}
\def\Z#1{_{\lower2pt\hbox{$\scriptstyle#1$}}}
\def\X#1{_{\lower2pt\hbox{$\scriptscriptstyle#1$}}}
\def\av{{a\ns{v}\hskip-2pt}} \def\aw{{a\ns{w}\hskip-2.4pt}}
\def\QQ{{\cal Q}} \def\qh{q}
\def\tw{\tc} \def\kv{k\ns v} \def\kw{k\ns w}
\def\Om{\Omega\Z M} \def\gw{\gb\ns w}
\def\fw{{f\ns w}} \def\fvi{{f\ns{vi}}} \def\fwi{{f\ns{wi}}}
\def\rhw{{\rh\ns w}} \def\rhv{\rh\ns v}
\def\etw{\eta\ns w} \def\etv{\eta\ns v} \def\rw{r\ns w}
\def\gv{\gb\ns v} \def\fv{{f\ns v}} \def\Hv{H\ns v} \def\Hw{H\ns w}
\def\FF{{\cal F}} \def\FI{\FF\Z I} \def\Fi{\FF\X I}
\def\OMM{\OM\Z M}\def\OMk{\OM\Z k}\def\OMQ{\OM\Z{\QQ}}\def\OMR{\OM\Z R}
\def\OmM{\Omega\Z M} \def\OmR{\Omega\Z R} \def\OMMn{\OM\Z{M0}}
\def\OmMw{\Omega\Z{M\hbox{\sevenrm w}}} \def\OmMn{\Omega\Z{M0}}
\def\fvf{\left(1-\fv\right)} \def\Hb{\bH\Z{\!0}} \def\Hh{H} \def\Hm{H\Z0}
\font\sevenrm=cmr7 \def\ws#1{_{\hbox{\sevenrm #1}}} \def\rs{r\ns s}
\def\ave#1{\langle{#1}\rangle} \def\Rav{\ave{\cal R}} \def\hr{h_r}
\def\gd{{{}^3\!g}} \def\pt{\partial} \def\noi{\noindent} \def\hri{h_{ri}}
\def\half{\frn12} \def\DD{{\cal D}} \def\bx{{\mathbf x}} \def\Vav{{\cal V}}
\def\frn#1#2{{\textstyle{#1\over#2}}} \def\VV{${\cal V}$}
\def\lsim{\mathop{\hbox{${\lower3.8pt\hbox{$<$}}\atop{\raise0.2pt\hbox{$\sim$}}
$}}} \def\dL{d\Z L} \def\dA{D\Z A} \def\rhcr{\rh\ws{cr}}
\def\kmsMpc{\w{km}\;\w{sec}^{-1}\w{Mpc}^{-1}} \def\ab{{\bar a}}
\def\epi{\epsilon_i}\def\gbi{\gb_i}\def\Omi{\OM_i}\def\OMkn{\OM_{k0}}
\def\fvn{f\ns{v0}} \def\Ci{C_\epsilon} \def\te{t_\epsilon}
\def\beq{\begin{equation}} \def\eeq{\end{equation}}
\def\bea{\begin{eqnarray}} \def\eea{\end{eqnarray}}
\def\PRL#1{Phys.\ Rev.\ Lett.\ {\bf#1}} \def\PR#1{Phys.\ Rev.\ {\bf#1}}
\def\ApJ#1{Astrophys.\ J.\ {\bf#1}} \def\CQG#1{Class.\ Quantum Grav.\ {\bf#1}}
\def\GRG#1{Gen.\ Relativ.\ Grav.\ {\bf#1}}

\title{Exact solution to the averaging problem in cosmology}

\author{David L. Wiltshire}
\affiliation{Department of Physics \& Astronomy, University of Canterbury,
Private Bag 4800, Christchurch 8140, New Zealand}

\begin{abstract}
The exact solution of a two--scale Buchert average of the Einstein equations
is derived for an inhomogeneous universe which represents a close
approximation to the observed universe. The two scales represent voids, and
the bubble walls surrounding them within which clusters of galaxies are
located. As described elsewhere [New J.\ Phys.\ {\bf9} (2007) 377],
apparent cosmic acceleration can be recognised as a
consequence of quasilocal gravitational energy gradients
between observers in bound systems and the volume average position in
freely expanding space. With this interpretation,
the new solution presented here replaces the Friedmann
solutions, in representing the average evolution of a
matter--dominated universe without exotic dark energy, while being
observationally viable.
\end{abstract}
\pacs{98.80.-k, 98.80.Es, 98.80.Jk, 95.36.+x}
\maketitle

At the time of last--scattering the distribution of matter in the
universe was very smooth, given the evidence of the cosmic microwave
background (CMB). At the present epoch, by contrast, the universe
is very lumpy on scales less than 100--300Mpc, with clusters of galaxies
strung in filaments and bubbles surrounding huge voids.
Some 40--50\% of the present volume \cite{HV} of the universe
is in voids of order $30h^{-1}$Mpc in diameter, $h$ being the dimensionless
Hubble parameter, $\Hm=100h\kmsMpc$, and when smaller and larger voids
\cite{morevoids} are taken into account, then the observable universe is
``void--dominated''.

In spite of present--day inhomogeneity, a broadly isotropic Hubble
flow is observed when one averages on sufficiently large scales.
This is taken as justification for assuming that cosmic
evolution can be modelled by the Friedmann equation for a smooth
fluid, despite the obvious observational evidence that galaxies are not
smoothly distributed. To achieve agreement with a number
of independent observations within the smooth fluid paradigm,
dark energy has been included in the standard cosmological model, posing
a foundational mystery for physics.

In recent years, a number of cosmologists have questioned whether the
observations, which have been interpreted as cosmic acceleration, might in
fact be accounted for by taking more care in deriving the geometry that
comes from averaging the actual inhomogeneous matter distribution.
In particular, the geometry which arises from the time evolution
of an initial average of the matter distribution does not generally
coincide, at a later time, with the average geometry of the full inhomogeneous
matter distribution evolved via Einstein's equations \cite{buch1}.
Whether or not the resulting ``back-reaction'' of inhomogeneities
on the average geometry can be large enough to explain effects usually
attributed to cosmic acceleration from dark energy in
Friedmann--Lema\^{\i}tre--Robertson--Walker (FLRW)
models, has been the subject of intense debate. (See
\cite{Crev} and references therein.)

In this {\em Letter} I will derive the general exact solution to the
Buchert equations \cite{buch1} for the two--scale model introduced in
ref.\ \cite{opus}, yielding a simple new observationally viable model of the
universe. The observational claim is based on my proposal that the debate
about dark energy from structure formation \cite{Crev} can be resolved by
careful consideration of the operational interpretation of measurements in
cosmology from first principles  \cite{opus}. This is necessary when averaging
an inhomogeneous cosmology, since in general the rods and clocks of observers
will be calibrated differently from those at an average location. In
writing down average parameters one must define how they are related to our
measurements operationally. Assuming the
Copernican principle, the fact that we observe an almost
isotropic CMB means that other observers should also measure an almost
isotropic CMB. However, it does not demand that such
observers measure the same mean CMB temperature as we, nor the same angular
scale for the Doppler peaks in the anisotropy spectrum. Significant
differences can arise due to gradients in spatial curvature
and associated gravitational energy.

In general relativity space is dynamical and can carry energy and momentum.
By the strong equivalence principle, since the laws of physics must coincide
with those of special relativity at a point, it is only internal energy that
can be localised in an energy--momentum tensor on the r.h.s.\ of the
Einstein equations. Thus the uniquely relativistic aspects of gravitational
energy associated with gradients in spatial curvature, and gradients in the
kinetic energy of spatial expansion, cannot be included in the energy
momentum tensor, but are at best described by a quasilocal formulation.
(For a review, see \cite{quasi}.)

In ref.\ \cite{opus} I propose a quantitative solution to the problem of
apparent cosmic acceleration through the technical definition of a
{\em finite infinity} scale, realising a qualitative suggestion of Ellis
 \cite{fit1}. Finite infinity replaces the usual notion of spatial infinity
in exact asymptotically flat spacetimes, as the fiducial reference point
for quasi--local gravitational energy with respect to observers in virialised
bound systems. A universal definition of this scale is possible, since the
initial expansion rate of the universe was extremely smooth at last
scattering, leading to a true critical density, $\rhcr$, as a demarcation
between potentially bound and unbound systems. Due to back--reaction,
$\rhcr$, does not evolve by the Friedmann equation.

While it has long been understood that averaging an inhomogeneous universe
entails the dressing of average cosmological parameters \cite{dress1}
through volume factors which relate to differences in spatial curvature,
the proposal of ref.\ \cite{opus} recognises that clock rates can also
vary systematically between observers in bound systems within finite
infinity, and a volume--average position in freely expanding space,
due to differences in gravitational energy. By this means an implicit
solution of the Sandage--de Vaucouleurs paradox \cite{opus} is possible:
the locally defined or bare Hubble parameter, $\bH$, can be uniform even
though voids appear to expand faster than the bubble walls which surround them,
since cosmic clocks within voids tick faster on account of gravitational
energy differences. Since our cosmological observations involve
photons exchanged with objects in bound systems, we do not
observe clocks in freely expanding space directly. Nonetheless,
an ideal comoving observer within a void would
measure a somewhat older age of the universe, and an isotropic CMB
with a lower mean temperature and an angular
anisotropy scale shifted to smaller angles.
\smallskip

Buchert's scheme is somewhat heuristic,
since it does not average all of the Einstein equations and requires
an extra integrability condition to ensure closure. However, starting
from a fully covariant averaging scheme \cite{Zal}, with reasonable
cosmological assumptions, the correlation tensor takes the form of a
spatial curvature \cite{CPZ}, and Buchert's scheme can be realised as
a consistent limit \cite{PS}. Following ref.\ \cite{opus}, the
Buchert average constructed here is based on the two scales most relevant to
the observed universe: (i) the voids which dominate the universe at
the present epoch; and (ii) finite infinity regions containing galaxy
clusters within the filaments and bubble walls that surround voids. The
local average geometry at the boundary of a finite infinity region is
assumed to be spatially flat, with the metric
\beq\ds^2\Z{\Fi}=-\dd\tw^2+\aw^2(\tw)\left[\dd\etw^2+
\etw^2\dd\Omega^2\right]\,.
\label{figeom}\eeq
Within voids the metric is not given by (\ref{figeom}) but is negatively
curved, with local scale factor $\av$ .
We average over the entire present epoch horizon volume,
$\Vav=\Vav\ns i\ab^3$, where
$\ab^3=\fvi\av^3+\fwi\aw^3$;
$\fvi$ and $\fwi=1-\fvi$
being the respective initial void and wall volume fractions at last
scattering, 
to construct the Buchert average geometry
\beq
\ds^2=-\dd\ts^2+\ab^2(\ts)\,\dd\etb^2+A(\etb,\ts)\,\dOM.
\label{avgeom}\eeq
Here the area function $A$ is defined by a horizon-volume average \cite{opus}.
The time--parameter $\ts$ differs from the wall--time $\tw$ of (\ref{figeom})
by the mean lapse function $\dd\ts=\gb(\tw)\,\dd\tw$. The geometry
(\ref{avgeom}) is not locally isometric to the local geometry in either the
walls or void centres.

When the geometry (\ref{figeom}) is related to the average
geometry (\ref{avgeom}) by conformal matching of radial null geodesics
it may be rewritten
\beq \ds^2\Z{\Fi}=
-\dd\tw^2+{\ab^2\over\gb^2}\left[\dd\etb^2+\rw^2(\etb,\tw)\,\dOM\right]
\label{wgeom}\eeq
where
$\rw\equiv\gb\fvf^{1/3}\fwi^{-1/3}\etw(\etb,\tw)$. Two
sets of cosmological parameters are relevant: those
relative to an ideal observer at the volume--average position in freely
expanding space using the metric (\ref{avgeom}), and conventional
dressed parameters using the metric (\ref{wgeom}). The conventional
metric (\ref{wgeom}) arises in our attempt to fit a single global metric
(\ref{figeom}) to the universe with the assumption that average spatial
curvature and local clock rates everywhere are identical to our own,
which is no longer true.

The volume--average matter, curvature and kinematic back-reaction parameters
are given by $\OMM=8\pi G\rhb\Z{M0}\ab\Z0^3/[3\bH^2\ab^3]$,
$\OMk=-\kv\fvi^{2/3}\fv^{1/3}/[\ab^2\bH^2]$, and
$\OMQ=-\dot\fv^2/[9\fv(1-\fv)\bH^2]$ respectively,
where the average curvature is due to the voids only, which are assumed
to have $\kv<0$, an overdot denotes a derivative w.r.t.\ volume--average
time, $t$, and $\bH\equiv\dot\ab/\ab$ is the volume--average or bare
Hubble parameter. It satisfies
\beq \bH=\fv\Hv+\fw\Hw,\eeq
where $\Hv\equiv\dot\av/\av$ and $\Hw\equiv\dot\aw/\aw$ are the
regional average expansion rates of voids and walls, {\em as measured by
volume--average clocks}. We define $\hr(t)\equiv\Hw/\Hv<1$. The
independent Buchert equations \cite{buch1}, including the integrability
condition that ensures their closure, are
\bea
&&\OMM+\OMk+\OMQ=1,\label{Beq1}\\
&&\ab^{-6}\pt_t\left(\OMQ\bH^2\ab^6\right)+\ab^{-2}\pt_t\left(\OMk\bH^2\ab^2
\right)=0\,.\label{Beq2}
\eea

Conventional dressed parameters defined with respect to the geometry
(\ref{wgeom}), relevant to ``wall observers'' such as ourselves, do not
satisfy a simple relation analogous to (\ref{Beq1}). The conventional
matter density parameter, $\OmM$, is expected to take numerical values
similar to those we infer in FLRW models. It differs from the bare
volume--average density parameter, $\OMM$, according to $\OmM=\gb^3\OMM$.
The mean lapse function is given by
\beq\gb=1+\hr^{-1}(1-\hr)f_v\,,\label{clocks2}\eeq
as a result of the requirement that the bare, or ``locally'' measured,
expansion rate is uniform. The dressed Hubble parameter that we measure
as wall observers, the global average over both walls and voids, is not $\bH$,
but
\beq
\Hh=\gb\bH-\Dts\gb=\gb\bH-\gb^{-1}\Dtc\gb\,.
\label{42}
\eeq


The Buchert equations (\ref{Beq1}),
(\ref{Beq2}) may be reduced to the pair of first order equations
\bea
\fvf{\dot\ab\over\ab}-\frn13\dot\fv=
\sqrt{\OMMn\Hb^2(1-\epi)\fvf{\ab\Z0^3\over\ab^3}}\,,\label{eqn4}\\
{\dot\ab\over\ab}+{\dot\fv\over3\fv}={\Hb\ab\Z0\over\fv^{1/3}\ab}
\sqrt{{\OMkn\over\fvn^{1/3}}+\OMMn\epi{\ab\Z0\over\fv^{1/3}\ab}}\,,
\label{eqn5}
\eea
where $\epi\ll1$ is an integration constant, obtained from a first integral
\cite{opus} of eqs.\ (\ref{Beq1}) and (\ref{Beq2}), namely
$$(1-\epi)\,\gb^2\OMM(1-\fv)^{-1}=1.$$
Eqs.\ (\ref{eqn4}) and (\ref{eqn5}) are readily integrated. Firstly
we multiply (\ref{eqn4}) by $\Hb^{-1}\fvf^{-2/3}\ab$ to obtain
$$ {1\over\Hb}\Der\dd t\left[\fvf^{1/3}\ab\right]=
\sqrt{\OMMn(1-\epi)\ab\Z0^3\over\fvf^{1/3}\ab}
$$
with the solution
\beq
\fvf^{1/3}\ab=\ab\Z0\left[(1-\epi)\,\OMMn\right]^{1/3}
\left(\frac32\Hb t\right)^{2/3}\,,
\label{sol1}\eeq
where a constant of integration corresponding to the origin of time has
been set to zero without loss of generality.
Since $\fwi^{1/3}\aw=\fvf^{1/3}\ab$, we see that
$\aw=a\ns{w0}t^{2/3}$,
where $a\ns{w0}\equiv\ab\Z0\left[\frac94\fwi^{-1}(1-\epi)\,\OMMn{\Hb}^2
\right]^{1/3}$. Thus the local expansion rate within the wall regions is
exactly that of an Einstein--de Sitter universe, {\em but in volume
average time not locally measured wall time}.

Finally, we multiply (\ref{eqn5}) by $\ab\Z0^{-1}\Hb^{-1}\fv^{1/3}\ab$ to
obtain
\beq {1\over\Hb}\Deriv\dd t u=
\sqrt{{\OMkn\over\fvn^{1/3}}\left(1+{\Ci\over u}\right)}\,,
\label{ode2}\eeq
where
$ u\equiv\fv^{1/3}\ab/\ab\Z0=\fvi^{1/3}\av/\ab\Z0$,
is proportional to $\av$, and $\Ci\equiv\epi\OMMn\fvn^{1/3}/\OMkn$
is a constant which can either be positive, zero or negative depending
on the initial value $\epi$.
Integrating (\ref{ode2}) we find
\beq
\sqrt{u(u+\Ci)}-\Ci\ln\left(\left|u\over\Ci\right|^{1\over2}
+\left|1+{u\over\Ci}\right|^{1\over2}\right)
={\al\over\ab\Z0}\left(t+\te\right)
\label{sol2}\eeq
where $\al=\ab\Z0\Hb\OMkn^{1/2}/\fvn^{1/6}$, and $\te$ is a constant which
cannot be chosen to be zero without loss of generality, as the time origin
was already fixed in determining (\ref{sol1}).

Eqs.\ (\ref{sol1}) and (\ref{sol2}) constitute the general exact solution
to the two--scale Buchert equations (\ref{Beq1}), (\ref{Beq2}), regardless
of the physical interpretation of observable quantities. In particular,
the Buchert equations have been studied by a number of authors without
mention as to what the physical interpretation of the time parameter, $t$,
is. Here we will pursue the interpretational framework of ref.\ \cite{opus}.

A number of quantities of interest may be found directly. Since
$\Hw=\dot\aw/\aw=2/(3t)$, and
$$\Hv={\dot\av\over\av}={2\over3t}\sqrt{{\fvf\epi\over\fv(1-\epi)}\left(
1+{u\over\Ci}\right)}$$
it follows that
\beq
\hr={\Hw\over\Hv}=\sqrt{(1-\epi)\OMMn\fvn^{1/3}\fv\over\left(\OMkn u+
\OMMn\fvn^{1/3}\epi\right)\fvf}\,.
\label{eqhr}\eeq
Evaluating [$\hbox{(\ref{eqn4})}+\fv\hbox{(\ref{eqn5})}$] at the present
epoch, we obtain the following constraint on parameters
\beq
\sqrt{(1-\epi)\OMMn(1-\fvn)}+\sqrt{(\OMkn+\OMMn\epi)\fvn}=1\,.
\label{con1}\eeq
This reduces the number of free parameters by one; we may take $\OMkn$ as
dependent, for example. The value of $\te$ is also
determined in terms of the other parameters by evaluating (\ref{sol2}) at
the present epoch, to give
\bea
&&{\OMkn^{3/2}\over\fvn^{1/2}}\Hb(\tn+\te)=\sqrt{\OMkn(\OMkn+\OMMn\epi)}
\nonumber\\ &&\quad -\OMMn\epi\ln\left[\sqrt{\left|\OMkn\over\OMMn\epi
\right|}+\sqrt{\left|1+{\OMkn\over\OMMn\epi}\right|}\,\right]\!,\label{con2}
\eea
where the age of the universe in volume--average time is
\beq
\tn={2\over3\Hb}\sqrt{1-\fvn\over(1-\epi)\OMMn}\,,
\eeq
on account of (\ref{sol1}). We observe that the
physical interpretation of the void volume--fraction ceases to be
physical meaningful in the limit $t\to0$ if $\te>0$, as is generally the
case. Radiation must be included to describe the universe at early times,
and has been omitted here. Here we restrict attention to the
matter--dominated epoch.

The general solution is specified by four independent
parameters, $\Hb$, $\epi$, $\OMMn$ and $\fvn$. However, two of these
may be further eliminated by taking priors \cite{paper2} at
the surface of last scattering consistent with the CMB.

Since eq.\ (\ref{sol2}) is a transcedental equation, the
combination of eqs.\ (\ref{sol1}) and (\ref{sol2}) only define $\ab(t)$
and $\fv(t)$ implicitly in the general case. Nonetheless, at late times
when both $t$ and $u$ are large, all general solutions to (\ref{sol2}) tend
to the particular solution with $\epi=0$ and $\te=0$. This particular
solution is in fact a late--time tracker solution, an attractor
which is insensitive to $\hri$ and $\fvi$, as long as the observable
universe is void--dominated.

As $\Ci=0$ when $\epi=0$, eq.\ (\ref{sol2}) yields a particularly
simple form for the late--time tracker solution. Since $\fvi^{1/3}\av=
\fv^{1/3}\ab =\ab\Z0 u$, we see that it
corresponds to the case in which the void regions expand exactly as
a Milne universe in volume--average time,
$\av=a\ns{v0}t$,
where $a\ns{v0}\equiv\OMkn^{1/2}\ab\Z0\Hb\fvn^{-1/6}\fvi^{-1/3}$.
It follows that $\hr=\Hw/\Hv=2/3$ is a constant for this special solution.
If we combine (\ref{con1}) with (\ref{eqhr}) evaluated at the present epoch,
we see that only one of the parameters $\OMMn$, $\OMkn$ and $\fvn$ is
independent, and in fact,
$\OMMn={4(1-\fvn)/(2+\fvn)^2}$,
$\OMkn={9\fvn/(2+\fvn)^2}$.

Consequently, the volume--average scale factor is
\beq
\ab={\ab\Z0\bigl(3\Hb t\bigr)^{2/3}\over2+\fvn}\left[3\fvn\Hb t+
(1-\fvn)(2+\fvn)\right]^{1/3}
\label{track1}\eeq
for the late--time tracker, while its void fraction is
\beq
\fv={3\fvn\Hb t\over3\fvn\Hb t+(1-\fvn)(2+\fvn)}\,,
\label{track2}\eeq
and the wall fraction is easily deduced from $\fw=1-\fv$. In terms of
$\fv(t)$, the mean
lapse function, bare Hubble parameter and bare matter density have the
simple forms $\gb(t)=\frn32 t\bH(t)=1+\half\fv$, and $\OMM(t)=
4(1-\fv)/(2+\fv)^2$.

The tracker--solution dressed Hubble parameter (\ref{42}) is
\beq
\Hh={2\over3\ts}+{\fv(4\fv+1)\over6t}\,,
\label{Htrack}\eeq 
Other dressed parameters \cite{opus} using the metric (\ref{wgeom}) relevant
to galactic wall obsevers include the redshift, $z$, and luminosity
distance, $\dL=\gc^{-1}\ab\Z0(1+z)\rw$, where $\rw=\gb\fvf^{1/3}
\int_\ts^{\ts\X0}\dd\tb/[\gb(\tb)(1-\fv(\tb))^{1/3}\ab(\tb)]$. For the tracker
solution these respectively satisfy
\bea
z+1&=&{\ab\Z0\gb\over\ab\gc}={(2+\fv)\fv^{1/3}\over3\fvn^{1/3}\Hb t}\,,
\\
{\Hb\dL\over(1+z)^2}&=&\left(\Hb\ts\right)^{2\over3}\int_\ts^{\ts\X0}
{2\Hb\dd\tb\over(2+\fv(\tb))(\Hb\tb)^{2/3}}\,.
\label{Dtrack}\eea
This last integral can be given in a simple closed analytic form.
These expressions are given in terms of volume--average time. To convert
to wall time relevant to galactic observers, we perform
the integral, $\tc=\int^\ts_0\dd\ts\,\gb^{-1}$, giving
\beq
\tc=\frn23\ts+{4\OmMn\over27\fvn\Hb}\ln\left(1+{9\fvn\Hb t
\over4\OmMn}\right)\,, \label{tsol}
\eeq
where $\OmMn=\frn12(1-\fvn)(2+\fvn)$ is the present epoch dressed matter
density in the case of the tracker solution. In (\ref{Htrack})--(\ref{Dtrack})
the parameter $t$ should be considered to be implicitly defined by (\ref{tsol})
in terms of wall time, $\tw$. At late times as $\ts\to\infty$, 
$\tc\goesas\frn23t$, so that $\Hh\goesas\frn32t^{-1}\goesas\tc^{-1}$, as
for an empty Milne universe. 

The age of the universe in volume--average time is $\tn=(2+\fvn)/(3\Hb)$, and
\beq
\tc\Z0={2(2+\fvn)\over9\Hb}\left[1+{(1-\fvn)\over3\fvn}\ln\left(2+\fvn\over
2(1-\fvn)\right)\right]\,,
\eeq
in wall time. The dressed Hubble constant we measure, $\Hm$, is
related to the bare Hubble constant, $\Hb$, by
\beq
\Hm={(4\fvn^2+\fvn+4)\Hb\over2(2+\fvn)}\,.
\eeq
Finally, the volume--average tracker--solution deceleration parameter
is $\bq=\half\OMM+2\OMQ=2\fvf^2/(2+\fv)^2$,
which begins close to the Einstein--de Sitter value, $\bq\goesas\half$,
when $\fv$ is small, and approaches $\bq\to0^+$ at late times, but remains
positive at all times. Thus a volume--average observer in freely
expanding space detects no cosmic acceleration. Nonetheless, a bound system
observer measures an effective dressed deceleration parameter
\beq
\qh={-\fvf(8\fv^3+39\fv^2-12\fv-8)\over\left(4+\fv+4\fv^2\right)^2}\,,
\label{qtrack}\eeq
which also begins at the Einstein--de Sitter value, $\qh\goesas\half$, for
small $\fv$ but then changes sign at epoch when $\fv\simeq0.5867$, at a zero
of the cubic in (\ref{qtrack}). Apparent acceleration reaches a maximum
when $\fv\simeq0.7736$ when $\qh\simeq-0.043$, and
then $\qh\to0^-$ at late times. Thus a wall observer registers a
late--time evolution which again is close to that of a Milne universe,
but this time with {\em apparent acceleration}.

While the initial conditions of the CMB require that $\hr\to1$ at
last scattering, for observationally
relevant initial conditions, the general solution (\ref{sol1}), (\ref{sol2})
approaches the tracker solution to within 1\% by a redshift of $z\goesas37$.
Thus the tracker solution can be used as a very reliable approximation all
the way back to the epoch of reionization. It is effectively the simplest
viable generalisation of the Einstein--de Sitter and Friedmann models, which
incorporates back--reaction, and is potentially of great utility.
Detailed cosmological
parameter fits are presented elsewhere \cite{paper2}, and demonstrate that the
present {\em fractal bubble model} \cite{opus} is at the very least a
serious contender for quantitatively solving the problem of dark energy
purely within general relativity. The analytic solution presented here,
and its simple tracker limit, should therefore provide the basis for many
future cosmological tests.

\vfill

\begin{thebibliography}{66}

\bibitem{HV}
F.~Hoyle and M.S.~Vogeley,
\ApJ{566}, 641 (2002); 
\ApJ{607}, 751 (2004). 

\bibitem{morevoids}
A.V.~Tikhonov and I.D.~Karachentsev,
\ApJ{653} (2006) 969; 
L.~Rudnick, S.~Brown and L.R.~Williams,
arXiv:0704.0908.

\bibitem{buch1}
T.~Buchert,
\GRG{32}, 105 (2000). 

\bibitem{Crev}
S.~R\"as\"anen,
JCAP {\bf 11}, 003 (2006); 
M.N.~C\'el\'erier,
astro-\break ph/0702416;
A.A.~Coley,
arXiv:0704.1734; 
T.~Buchert,
AIP Conf.\ Proc.\  {\bf 910}, 361 (2007); 
arXiv:0707.2153. 

\bibitem{opus}
D.L.~Wiltshire,
New J.\ Phys.\ {\bf9}, 377 (2007). 

\bibitem{quasi}
L.B.~Szabados,
Living Rev.\ Rel.\ {\bf 7}, 4 (2004).

\bibitem{fit1}
G.F.R.~Ellis,
in B.~Bertotti, F.~de Felice and A.~Pascolini (eds), {\it General
Relativity and Gravitation}, (Reidel, Dordrecht, 1984) pp.~215--288.

\bibitem{dress1}
T.~Buchert and M.~Carfora,
\PRL{90}, 031101 (2003); 
\CQG{19}, 6109 (2002). 

\bibitem{Zal}
R.M.~Zalaletdinov,
\GRG{24}, 1015 (1992);
{\bf25}, 673 (1993).

\bibitem{CPZ}
A.A.~Coley, N.~Pelavas and R.M.~Zalaletdinov,
Phys.\ Rev.\ Lett.\ {\bf 95} (2005) 151102;
A.A.~Coley and N.~Pelavas,
\PR{D 74}, 087301 (2006); 
{\bf D 75}, 043506 (2007). 

\bibitem{PS} 
A.~Paranjape and T.P.~Singh,
\PR{D 76}, 044006 (2007).  

\bibitem{paper2}
B.M.~Leith, S.C.C.~Ng and D.L.~Wiltshire,
arxiv:0709.2535.
\end{thebibliography}
\end{document}